\documentclass[12pt,preprint]{aastex}
\usepackage{amsmath,amssymb}
\usepackage{graphicx}
\usepackage{float}
\usepackage{multirow}
\usepackage{pdflscape}

\begin{document}

\title{Non-Gaussian Error Distributions of LMC Distance Moduli Measurements}

\author{Sara Crandall and Bharat Ratra}

\affil{Department of Physics, Kansas State University, 
                 116 Cardwell Hall, Manhattan, KS 66506, USA} 
\email{sara1990@phys.ksu.edu and ratra@phys.ksu.edu}

\begin{abstract}
We construct error distributions for a compilation of 232 Large Magellanic Cloud (LMC) distance moduli values from \cite{Grijs2014} that give an LMC distance modulus of $(m-M)_{0}=18.49 \pm 0.13$ mag (median and $1\sigma$ symmetrized error). Central estimates found from weighted mean and median statistics are used to construct the error distributions. The weighted mean error distribution is non-Gaussian --- flatter and broader than Gaussian --- with more (less) probability in the tails (center) than is predicted by a Gaussian distribution; this could be the consequence of unaccounted-for systematic uncertainties. The median statistics error distribution, which does not make use of the individual measurement errors, is also non-Gaussian --- more peaked than Gaussian --- with less (more) probability in the tails (center) than is predicted by a Gaussian distribution; this could be the consequence of publication bias and/or the non-independence of the measurements. We also construct the error distributions of 247 SMC distance moduli values from \cite{Grijs2015}. We find a central estimate of $(m-M)_{0}=18.94 \pm 0.14$ mag (median and $1\sigma$ symmetrized error), and similar probabilities for the error distributions.

\end{abstract}

\maketitle
\section{Introduction}
The LMC is a widely studied nearby extragalactic setting with a plethora of stellar tracers. The closeness of the LMC and the abundance of tracers has resulted in a large number of distance measurements to this nearby galaxy. As the LMC distance provides an important low rung of the cosmological distance ladder, it is of great interest to study collections of LMC distance moduli measurements. Following \cite{Schaefer2008}, \cite{Grijs2014} compiled a list of 237 LMC distance moduli published during 1990-2014\footnote{Five of the \cite{Grijs2014} entries do not have error bars, so here we only consider the 232 measurements that do.} and used these data to examine the effects of publication bias and correlation between the measurements. They conclude that the overall effect of publication bias is not strong, although there are significant effects due to measurement correlations, especially in some individual tracer (smaller) subsamples. 

In this paper we extend and complement the analysis of \cite{Grijs2014} by constructing and studying the error distributions of the full (232 measurements) sample and two individual tracer subsamples of the \cite{Grijs2014} compilation. More specifically, we examine the Gaussianity of these error distributions.\footnote{Conventionally one assumes a Gaussian distribution of errors. For instance, this is used when determining constraints from CMB anisotropy data \citep[see e.g.,][]{Ganga1997,Ratra1999,Chen2004,Bennett2013} and has been tested for such data \citep[see e.g.,][]{Park2001,Ade2015}. \cite{Schaefer2008} also assumes the LMC distance moduli measurement errors have a Gaussian distribution.} We begin by following \cite{Chen2003b} and \cite{Crandall2015} and construct an error distribution, a histogram of measurements as a function of $N_{\sigma}$, the number of standard deviations that a measurement deviates from a central estimate. This is similar to the z score analysis of \cite{Grijs2014}, however, we use a central estimate from the data compilation itself whereas \cite{Grijs2014} use two published values that are assumed to well represent the measurements. We use two techniques to find the central estimate: weighted mean and median statistics. Since median statistics does not make use of individual measurement error bars, median statistics constraints are typically weaker than weighted mean ones, but are more reliable in the presence of unaccounted-for systematic errors.

We find larger probability tails (error) in the weighted mean distributions. For the median 232 measurements case, we find that the distribution is narrower than a Gaussian distribution at small (intermediate) $N_\sigma\leq2$ where the probability is higher (lower) than expected for a Gaussian distribution (a similar effect is seen in the smaller sub-samples we study). We attempt to analytically categorize these distributions by fitting to well-known non-Gaussian distributions: Cauchy, Student's $t$, and the double exponential. Using a Kolmogorov-Smirnov (KS) test, the fits are poor ($<0.1\%$) for the Cauchy, and double exponential cases. A Student's $t$ case with a $n=39$ gives a probability of 21$\%$, and is the best fit. 

Given that the weighted mean error distributions are significantly non-Gaussian, it is proper to focus more on our median statistics central estimate results. In this case, for all three data sets, the error distributions are narrower than Gaussian. This could be the result of mild publication bias, or more likely, as argued by \cite{Grijs2014}, the consequence of correlations between measurements. 

In Section \ref{Methods} we summarize our methods of graphically and numerically describing the error distribution of the distance moduli values. Section \ref{All} describes our findings from analyses of the distribution of all 232 measurements. Sections \ref{Cepheids} and \ref{Lyrae} summarize our analyses of the two individual tracer subsamples. In Section \ref{SMC} we describe the results found using SMC distance moduli measurements from \cite{Grijs2015}. We conclude in Section \ref{Conclusion}.

\section{Summary of Methods}
\label{Methods}

Of the 237 LMC distance moduli values collected by \cite{Grijs2014}, five do not have a quoted error. For our analyses here we use the 232 measurements with symmetric statistical error bars. To determine the error distribution of the 232 measurements we must first find a central estimate. We do this using two statistical techniques: weighted mean and median statistics. 

The weighted mean \citep{Podariu01} is
\begin{equation}
{{D}}_{\mathrm{wm}}={\frac{\sum\limits_{i=1}^{N}{{D}}_{i}/\sigma_{i}^2}{\sum\limits_{i=1}^{N}1/\sigma_{i}^2}},
\end{equation}
where ${D}_{i}$ is the distance modulus and $\sigma_{i}$ is the one standard deviation error of $i=1,2,....,N$ measurements. While \cite{Grijs2014} use only the quoted statistical error, in our analyses $\sigma_{i}$ is the quadrature sum of the systematic (if quoted) and statistical errors. Since many do not quote a systematic error,\footnote{\cite{Grijs2014} note that only 49 measurements have a quoted non-zero systematic error, and four additional measurements include systematic uncertainties in their error. The significance of this is considered in Section \ref{Conclusion}.} and if one is stated it is small, the difference is not large. The weighted mean standard deviation is
\begin{equation}
\sigma_{\mathrm{wm}}=\left(\sum_{i=1}^{N}1/\sigma_{i}^{2}\right)^{-1/2}.
\end{equation}
We can also determine a goodness of fit, $\chi^{2}$, by
\begin{equation}
\chi^{2}=\frac{1}{N-1}\sum_{i=1}^{N}\frac{({{D}}_{i}-{D}_{\mathrm{wm}})^{2}}{\sigma_{i}^{2}}.
\end{equation}
The number of standard deviations that $\chi$ deviates from unity is a measure of good-fit and is given by

\begin{equation}
N_{\sigma}=|\chi-1|\sqrt{2(N-1)}.
\end{equation}

The median statistics technique is beneficial because it does not make use of the individual measurement errors. However, consequently, this will result in a larger uncertainty on the central estimate than for the weighted mean case. To use median statistics we assume that all measurements are statistically independent and have no systematic error as a whole. A measurement then has a $50\%$ chance of either being below or above the median value. For a detailed description of median statistics see \cite{Gott2001}.\footnote{For applications and discussions of median statistics see \cite{Chen2003a}, \cite{Mamajek2008}, \cite{Chen2011}, \cite{Calabrese2012}, \cite{Croft2015}, \cite{Andreon2012}, \cite{Farooq2013}, \cite{Crandall2014}, \cite{Ding2015}, \cite{Colley2015}, and \cite{Sereno2015}.} 

Once a central estimate is found, we can construct an error distribution using $N_{\sigma}$ defined as
\begin{equation}
\label{Nsig}
N_{\sigma_{i}}=\frac{D_{i}-D_{\rm{CE}}}{(\sigma_{i}^{2}+\sigma_{\rm{CE}}^{2})^{1/2}}
\end{equation}
where $D_{\rm{CE}}$ is the central estimate of $D_{i}$, either $D_{\mathrm{wm}}$ or $D_{\mathrm{med}}$, and $\sigma_{\rm{CE}}$ is the error of the central estimate, either $\sigma_{\rm{wm}}$ or $\sigma_{\rm{med}}$. Here $D_{\rm{med}}$ is the median distance modulus, with $50\%$ of the measurements being above it and $50\%$ below, and $\sigma_{\rm{med}}$ is defined as in \cite{Gott2001} such that the range $D_{\rm{med}} \pm \sigma_{\rm{med}}$ includes $68.3\%$ of the probability. \cite{Grijs2014} consider a similar variable, a ``z score". Their z score is different in that they use two reference values for their central estimate (\cite{Freedman2001} and \cite{Pietrzynski2013}) while we use the weighted mean and median central estimates. \cite{Grijs2014} assume that the reference values are a good representation of the distance moduli measurements. Therefore they use the z score assuming Gaussianity. We do not assume Gaussianity as our central estimates are found directly from the collected \cite{Grijs2014} data using our statistical techniques. 

To numerically describe the error distribution, we use a nonparametric Kolmogorov-Smirnov (KS) analysis. This is used to test the compatibility of a sample distribution to a reference distribution. This can be used in two ways, with binned or un-binned data.\footnote{It is more conventional to use un-binned data for this test, but for completeness we have used both (see Sec. 5.3.1 \cite{Feigelson2012}).} The test compares the LMC distance moduli measurements to a well know distribution function.

To set conventions, we first use the Gaussian probability distribution function is
\begin{equation}
P(|\textbf{x}|)=\frac{1}{\sqrt{2\pi}}\exp(-|\textbf{x}|^{2}/2).
\end{equation}
It will also be of interest to consider other well-known distributions. These include the Cauchy (or Lorentzian) distribution 
\begin{equation}
 P(|\textbf{x}|)=\frac{1}{\pi}\frac{1}{1+|\textbf{x}|^{2}}.
 \end{equation}
This distribution has extended tails, and is a popular choice for a widened distribution compared to the Gaussian. The Cauchy distribution has large extended tails with an expected $68.3\%$ and $95.4\%$ of the values falling within $|\textbf{x}|\leq1.8$ and $|\textbf{x}|\leq14$ respectively. The Student's $t$ distribution is described by
\begin{equation}
P(|\textbf{x}|)=\frac{\Gamma[(n+1)/2]}{\sqrt{\pi n}\Gamma(n/2)}\frac{1}{(1+|\textbf{x}|^{2}/n)^{(n+1)/2}}.
\end{equation}
Here $n$ is a positive parameter,\footnote{The inclusion of $n$ reduces the number of degrees of freedom by 1.} and $\Gamma$ is the gamma function. When $n\rightarrow\infty$ this becomes the Gaussian distribution. When $n=1$ it is the Cauchy distribution. Thus, for $n>1$, it is a function with extended tails, but less so than that of the Cauchy distribution. The last distribution that we consider is the double exponential. This is given by
\begin{equation}
 P(|\textbf{x}|)=\frac{1}{2}\exp{(-|\textbf{x}|)}.
 \end{equation}
The double exponential falls off less rapidly than a Gaussian distribution, but faster than a Cauchy distribution. For this distribution $68.3\%$ and $95.4\%$ of the values fall within $|\textbf{x}|\leq1.2$ and $|\textbf{x}|\leq3.1$ respectively. 

The comparison between the sample and assumed distribution yields a p-value (or probability) that the two are of the same distribution.

\section{Error Distribution for Full Dataset}
\label{All}

When using weighted mean statistics, the 232 LMC distance moduli yield a central estimate of  $(m-M)_{0}=18.49 \pm 3.11\times 10^{-3}$ mag. We also find $\chi^{2}=3.00$ and the number of standard deviations that $\chi$ deviates from unity is $N=15.7$. For the median case we find a central estimate of $(m-M)_{0}= 18.49$ mag with a $1\sigma$ range of $18.32$ mag $\leq (m-M)_{0}\leq18.59$ mag. Our central estimates are in good accord with \cite{Grijs2014} who quote $(m-M)_{0}=18.49 \pm 0.09$ mag.\footnote{\cite{Grijs2014} use a collection of 233 distance moduli values for their estimate from years 1990 to 2013, dropping four 2014 measurements.}

\begin{center}
\begin{figure}[H]
\advance\leftskip-1.25cm
\advance\rightskip-1.25cm
\includegraphics[height=68mm,width=95mm]{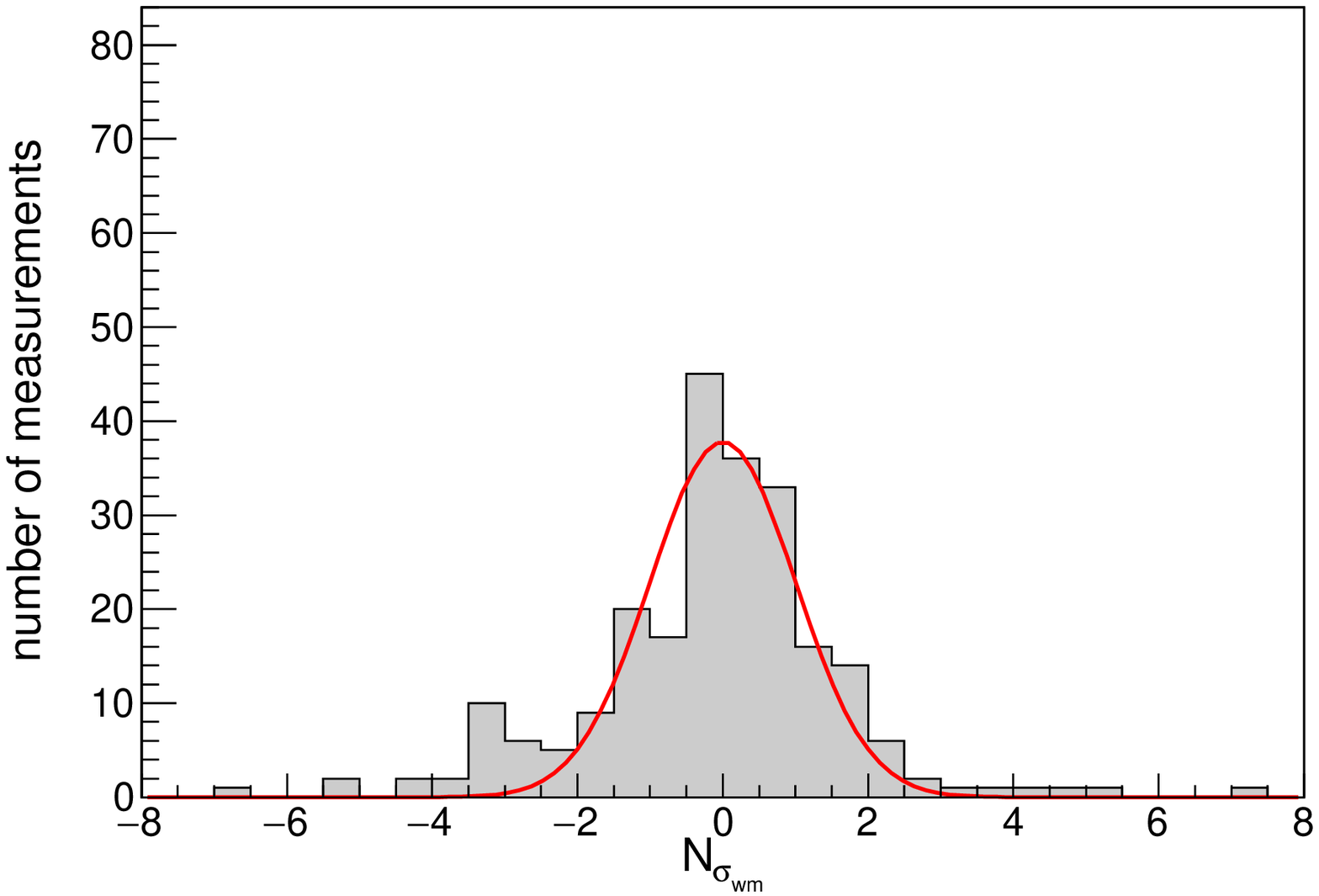}
\includegraphics[height=68mm,width=95mm]{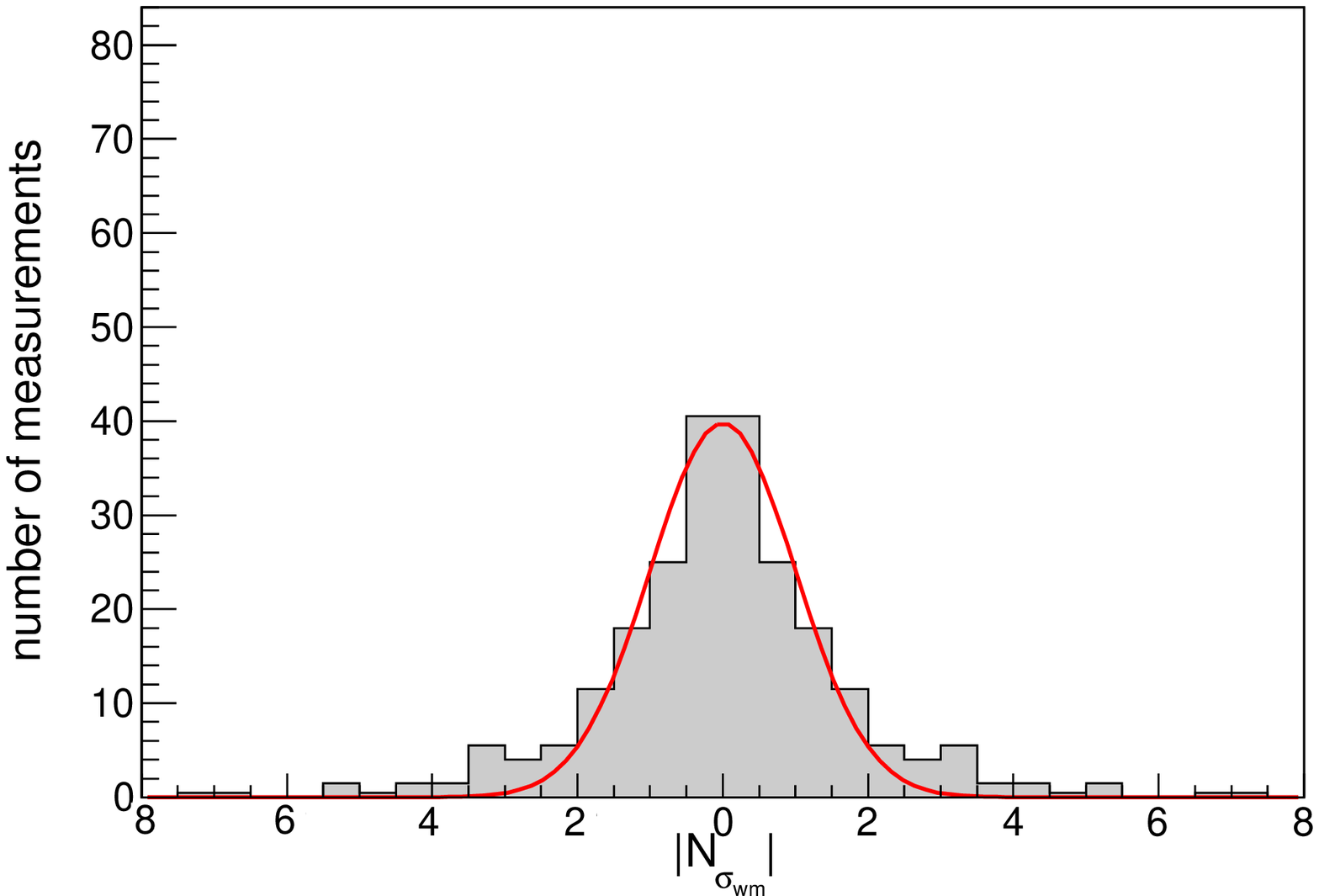}
\includegraphics[height=68mm,width=95mm]{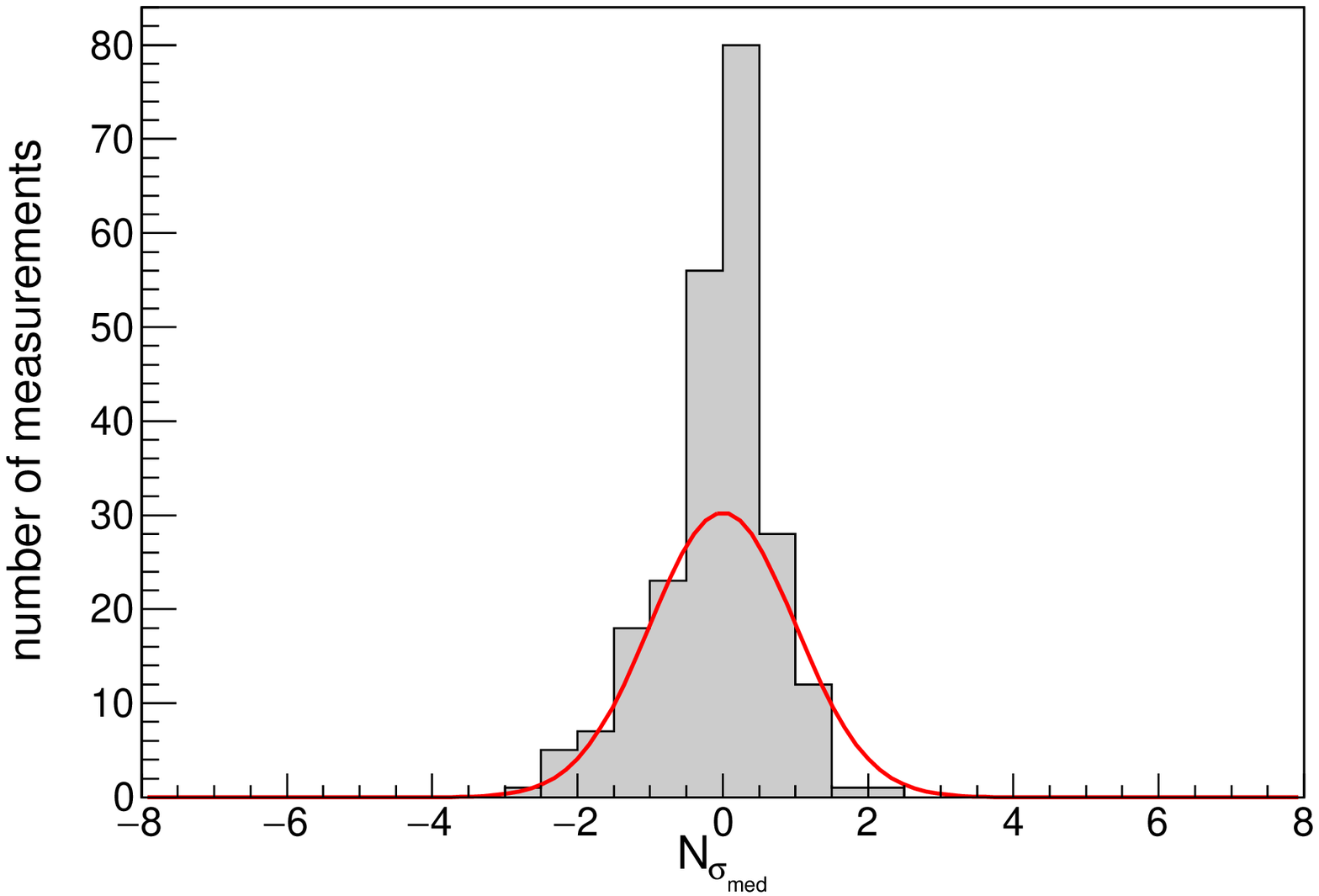}
\includegraphics[height=68mm,width=95mm]{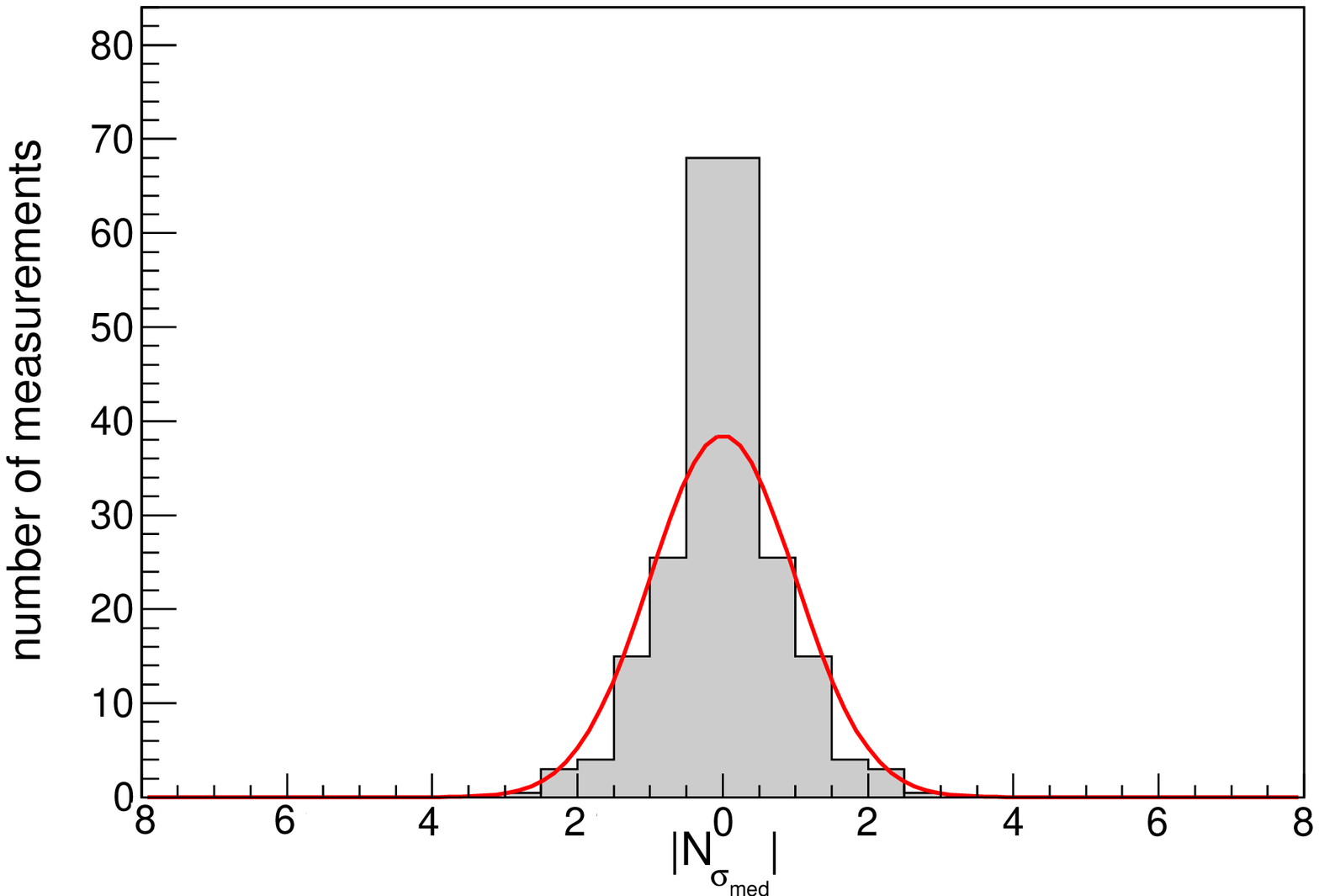}
\caption{Histograms of the error distribution in half standard deviation bins. The top (bottom) row uses the weighted mean (median) of the 232 measurements as the central estimate. The left column shows the signed deviation, where positive (negative) $N_{\sigma}$ represent a value that is greater (less) than the central estimate. The right column shows the absolute symmetrized distributions. The smooth curve in each panel is the best-fit Gaussian.} 
\label{figure:Nsig}
\end{figure}
\end{center}

\begin{figure}[H]
\includegraphics[width=\linewidth]{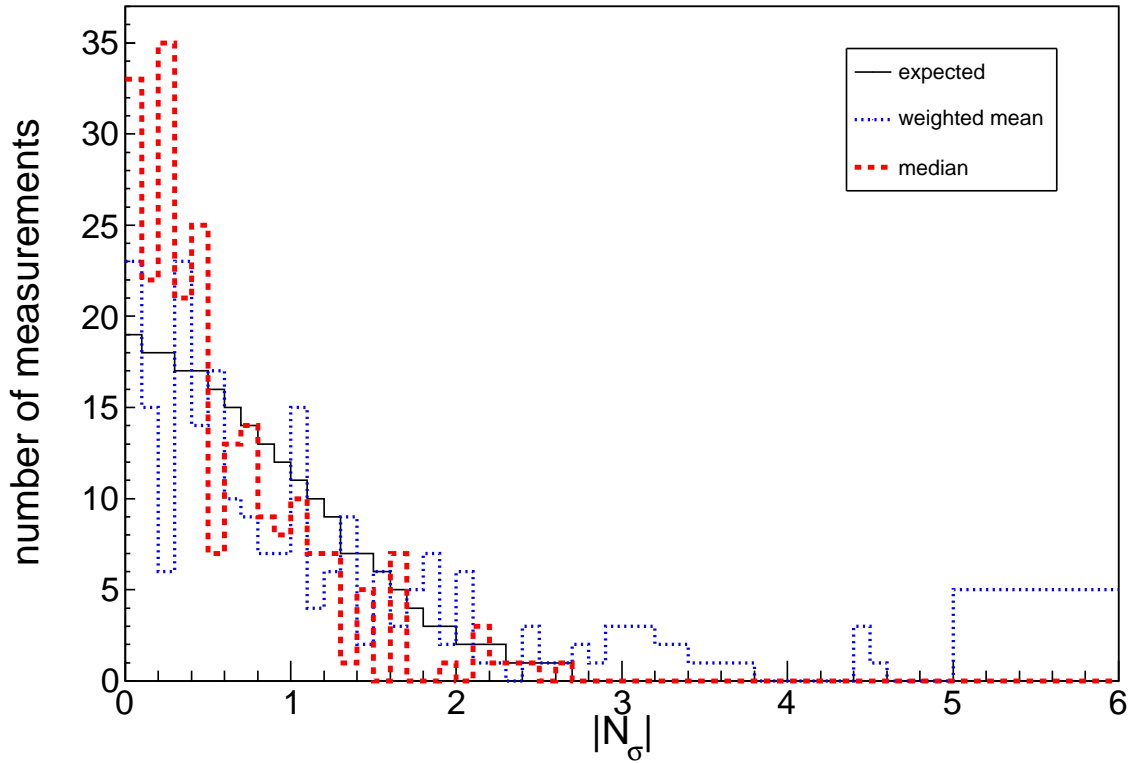}
\caption{Histogram of the error distributions in $|N_{\sigma}|=0.1$ bins (with the exception of the last, truncated, bin with $5\leq|N_{\sigma}|\leq6$ that contains the number of measurements to $|N_{\sigma}|=8$). The solid black line represents the expected Gaussian probabilities for 232 measurements and the dotted blue (dashed red) line is the number of $|N_{\sigma}|$ values in each bin for the weighted mean (median) case.} 
\label{figure:prob}
\end{figure}

Figure \ref{figure:Nsig} shows the error distribution of the 232 measurements. These are shown as a function of $N_{\sigma}$, Eq.\ \ref{Nsig}, the number of standard deviations the measured value deviates from the central estimate. In Fig.\ \ref{figure:Nsig} we show the error distributions for the weighted mean and median central estimates.\footnote{The larger $\sigma_{CE}$ for the median case results in a narrower distribution, see Eq.\ \ref{Nsig}.} In both cases we also plot the symmetrized distribution as a function of $|N_{\sigma}|$. For a more detailed perspective of the distribution, see Fig.\ \ref{figure:prob} (with $|N_{\sigma}|=0.1$ bin size). 

Figure \ref{figure:Nsig}  shows that for the weighted mean case the distribution has a more extended tail than expected for a Gaussian distribution. In fact, for a set of 232 values, a Gaussian distribution should yield 11 values with $|N_{\sigma}|\geq2$, one value with $|N_{\sigma}|\geq3$, and none with $|N_{\sigma}|\geq4$. However, we find 42 values with $|N_{\sigma}|\geq2$, 23 with $|N_{\sigma}|\geq3$ and nine with $|N_{\sigma}|\geq4$ for the weighted mean case. We also note that $68.3\%$ of the observed weighted mean $N_{\sigma}$ error distribution falls within $-1.37\leq N_{\sigma} \leq 1.26$ while $95.4\%$ lies within $-3.37\leq N_{\sigma} \leq 4.57$. The observed weighted mean $|N_{\sigma}|$ error distribution has limits of $|N_{\sigma}|\leq 1.33$ and $|N_{\sigma}|\leq 3.63$ respectively, and $56.5\%$ and $81.9\%$ of the values fall within $|N_{\sigma}|\leq 1$ and $|N_{\sigma}|\leq 2$ respectively. These results clearly indicate that the weighted mean error distribution is non-Gaussian and so the weighted mean technique is inappropriate for an analysis of these data.

The median case is narrower than Gaussian, with seven values of $|N_{\sigma}|\geq2$ and none with $|N_{\sigma}|\geq3$. $68.3\%$ of the data falls within $-0.86\leq N_{\sigma} \leq 0.63$ while $95.4\%$ lies within $-1.97\leq N_{\sigma} \leq 1.27$. The $|N_{\sigma}|$ error distribution has limits of $|N_{\sigma}|\leq 0.72$ and $|N_{\sigma}|\leq 1.66$ respectively, and $80.6\%$ and $97.0\%$ of the values fall within $|N_{\sigma}|\leq 1$ and $|N_{\sigma}|\leq 2$ respectively. The median technique is more appropriate because of the non-Gaussianity of the distributions, however, \cite{Grijs2014} note that there are correlations between measurements (especially among measurements of the same tracer type). These correlations mean that the measurements are not statistically independent, and the errors associated with the median will need to be slightly adjusted to account for this. Regardless, the narrowness of the median distribution is clearly consistent with the presence of such correlations. 

Since the distribution for the weighted mean case is broader than Gaussian while the median distribution is narrower than a Gaussian, it is of interest to try to fit these observed distributions to well-known non-Gaussian distributions.  

To set conventions, we first consider a Gaussian probability distribution function. In this case $68.3\%$ of the values have $|N_{\sigma}|\leq1$. The Gaussianity of the distribution can be established by taking a quantitative look at the spread of values. However, the probability given by the KS test is $\leq0.1$\% for the data set (See Table \ref{table:KS}). Our first non-Gaussian distribution, the Cauchy (or Lorentzian) distribution, also has a probability of $<0.1\%$.

Next we consider a distribution with extended tails, but less so than the Cauchy distribution, the Student's $t$ distribution. Fitting to this function yields a probability of $21\%$ (corresponding to a Student's $t$ distribution with $n=39$) for a binned KS test. This may appear odd, as we have argued that for the median case the distribution is narrower than a Gaussian distribution, while the Student's $t$ distribution is known for extended tails. To explain this we examine the kurtosis of the $|N_{\sigma}|$ distribution. We use the common definition of kurtosis (see Eq.\ 37.8b of \citealt{Olive2014})
\begin{equation}
k=\frac{m_{4}}{m_{2}^{2}},
\end{equation}
where the fourth and second moments are
\begin{equation}
m_{4}=\frac{1}{n}\sum\limits_{i=1}^{n}{(|N_{\sigma_{i}}|-\bar{|N_{\sigma}|})^{4}}
\end{equation} 
and
\begin{equation}
m_{2}=\frac{1}{n}\sum\limits_{i=1}^{n}{(|N_{\sigma_{i}}|-\bar{|N_{\sigma}|})^{2}}.
\end{equation}
Here $\bar{|N_{\sigma}|}$ is the mean of the $n$ $|N_{\sigma_{i}}|$ values. For a detailed discussion of kurtosis see \cite{Balanda1988}. The kurtosis can be defined as a measurement of the peakedness, or that of the tail width of a distribution. For example, a large kurtosis would represent a distribution with more probability in the peak and tails than in the ``shoulder'' \citep{Balanda1988}. A Gaussian distribution has $k=3$, and $k\geq3$ represents a leptokurtic distribution with a high peak and wide tail.\footnote{Often an ``excess kurtosis" is used to describe the peakedness of a distribution. This is simply three subtracted from the standard kurtosis, and is used to compare to a normal distribution (which would have an excess kurtosis of zero).}  For the median case, we find $k=5.57$. This may explain why this case favors a Student's $t$ fit, as a Student's $t$ distribution also has a large kurtosis, i.e. wider tails and a higher peak. The median statistics distribution appears to favor this fit because its kurtosis is greater than that for a Gaussian distribution, even though it is narrower than a Gaussian.

\begin{deluxetable}{llcc} 
\tablecaption{K-S Test Probabilities}
\tablewidth{0pt}
\tabletypesize{\small}

\tablehead{
&&Un-binned&Binned\\
\colhead{Function\tablenotemark{a}}&
\colhead{Data Set}&
\colhead{Probability(\%)\tablenotemark{b}}&
\colhead{Probability(\%)\tablenotemark{b}}\\
\noalign{\vskip -6mm}
}
\startdata
\noalign{\vskip -0mm}
\noalign{\vskip 0mm}
Gaussian			&	Whole (232)&				$<0.1$&		$<0.1$\\
				&	Truncated (223)&			$<0.1$&		$<0.1$\\
				&	Cepheids (81)&				1.5&			$<0.1$\\
				&	Truncated Cepheids (75)&	2.8&			0.10\\
				&	RR Lyrae (63)&				1.5&			$<0.1$\\
				&	Truncated RR Lyrae (58)&		0.8&			$<0.1$\\
Cauchy			&	Whole (232)&				$<0.1$&		$<0.1$\\
				&	Truncated (223)&			$<0.1$&		$<0.1$\\
				&	Cepheids (81)&				1.0&			$<0.1$\\
				&	Truncated Cepheids (75)&	2.9&			$<0.1$\\
				&	RR Lyrae (63)&				1.6&			$<0.1$\\
				&	Truncated RR Lyrae (58)&		0.7&			$<0.1$\\
Double Exponential 	&	Whole (232)&				$<0.1$&		$<0.1$\\
				&	Truncated (223)&			$<0.1$&		$<0.1$\\
				&	Cepheids (81)&				1.5&			$<0.1$\\
				&	Truncated Cepheids (75)&	3.7&			$<0.1$\\
				&	RR Lyrae (63)&				1.3&			$<0.1$\\
				&	Truncated RR Lyrae (58)&		0.6&			$<0.1$\\
$n=39$ Student's $t$&	Whole (232)&				$<0.1$&		21\\
$n=13$ Student's $t$&	Truncated (223)&			$<0.1$&		28\\
$n=3$ Student's $t$	&	Cepheids (81)&				0.9&			26\\
$n=22$ Student's $t$&	Truncated Cepheids (75)&	2.7&			25\\				
$n=59$ Student's $t$&	RR Lyrae (63)&				1.6&			34\\	
$n=94$ Student's $t$&	Truncated RR Lyrae (58)&		0.6&			37\\
\enddata
\tablenotetext{a}{For the Student's $t$ case, the $n$ corresponding with the best probability is displayed.}
\tablenotetext{b}{The probability that the data set is compatible with the assumed distribution.}
\label{table:KS}
\end{deluxetable}

The final non-Gaussian distribution function we consider is the double exponential, or Laplace distribution. Again, we do not find a probability of greater than $0.1\%$. 
 
To visually clarify the difference between the weighted mean and median statistics $|N_{\sigma}|$ histograms, we plot them in bins of $|N_{\sigma}|=0.1$, see Fig.\ \ref{figure:prob}. We see that the weighted mean case is closer to Gaussian near the peak, but has an extended tail. This suggests the existence of unaccounted-for systematic errors. For the median case the peak is much higher than expected for a Gaussian, and the distribution drops off with increasing $|N_{\sigma}|$ more rapidly than expected for a Gaussian. This may be a sign of correlations between measurements or possibly publication bias.

\begin{center}
\begin{figure}[H]
\advance\leftskip-1.25cm
\advance\rightskip-1.25cm
\includegraphics[height=68mm,width=95mm]{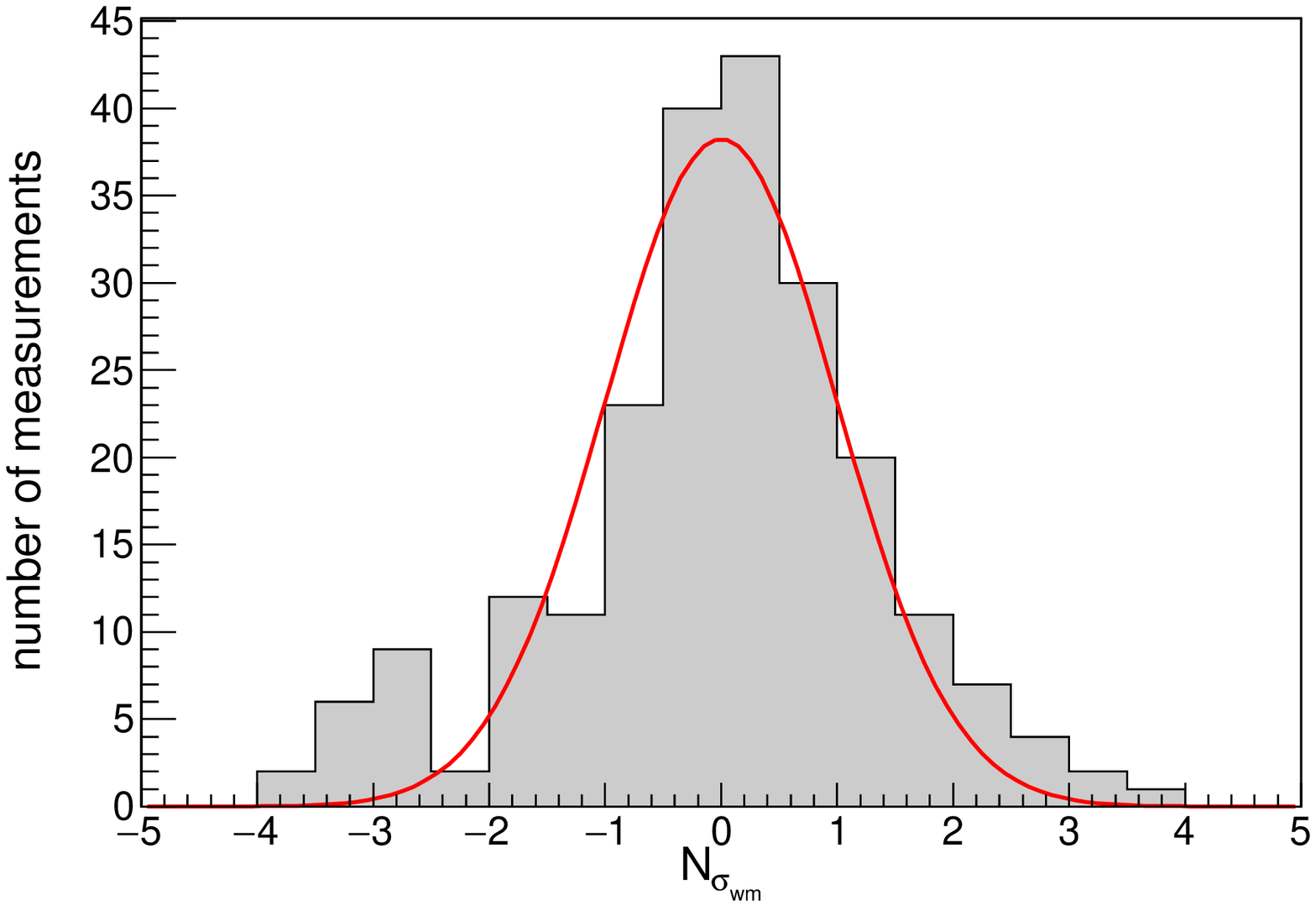}
\includegraphics[height=68mm,width=95mm]{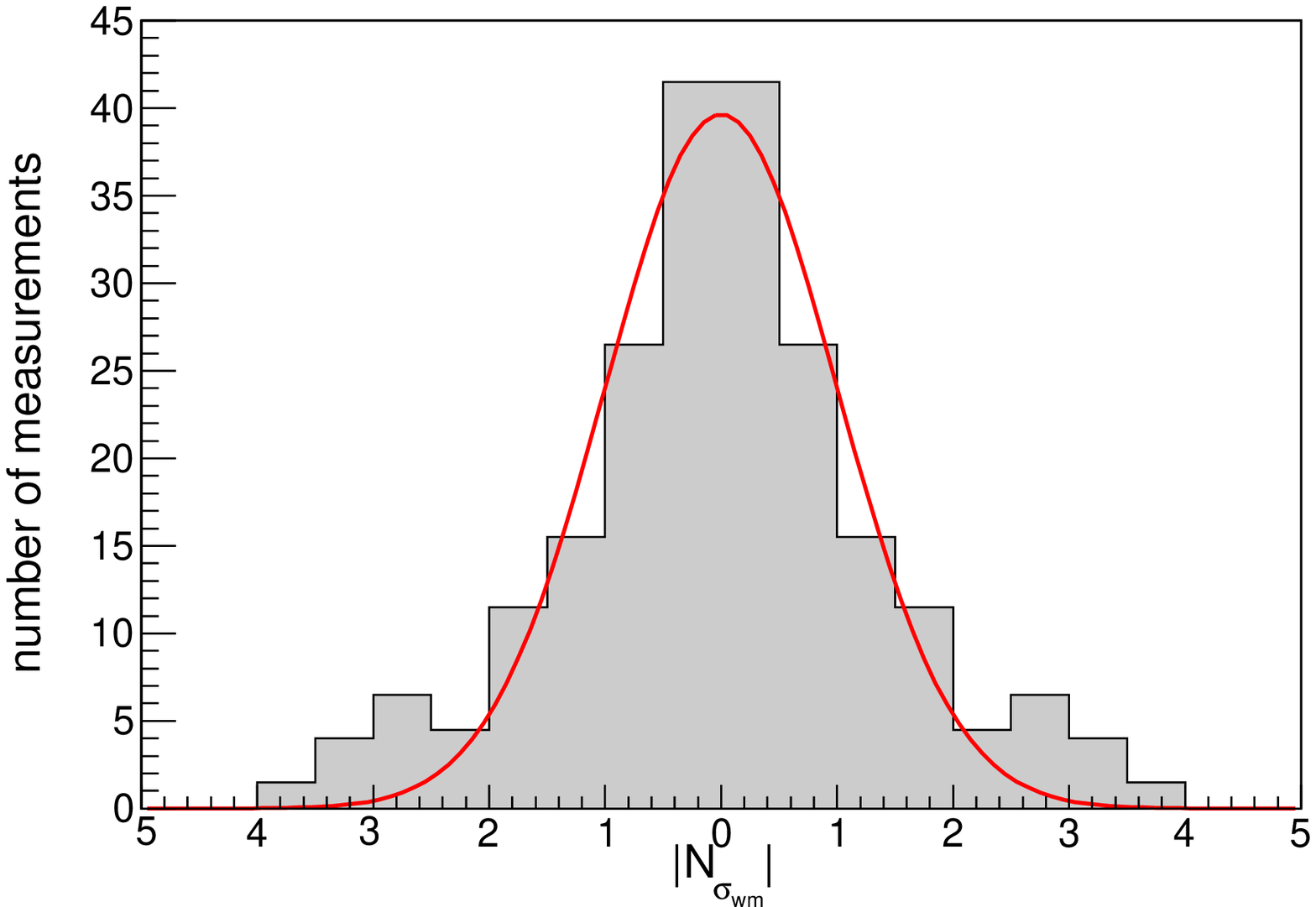}
\caption{Histograms of the error distribution in half standard deviation bins for the truncated weighted mean distribution ($N_{\sigma}\leq4$). The left plot uses the weighted mean of the 223 measurements as the central estimate to show the signed deviation. The right plot shows the symmetrized absolute $N_{\sigma}$. The smooth curve in each panel is the best-fit Gaussian.} 
\label{figure:Nsig cap}
\end{figure}
\end{center}

Table \ref{table:expected events Gaussian} is a more compact way of displaying some of this information. For a Gaussian distribution of 232 values, there should be zero measurements with $|N_{\sigma}|\geq4$ while the observed weighted mean case has nine. For illustrative purposes, we truncate this distribution by removing all values with $|N_{\sigma}|\geq4$.\footnote{For completeness, we also did a median statistics analysis of this truncated data. As expected, we found that removing these nine measurements does not increase probabilities or change the median statistics results, which shows the robustness of median statistics.} This leaves us with 223 values and an unchanged central estimate of  $(m-M)_{0}=18.49 \pm 3.38\times 10^{-3}$.\footnote{The 223 values also give a $\chi^{2}=1.90$ and $N=8.01$ (the number of standard deviations that $\chi$ deviates from unity).} The spread of the values can be seen in Fig.\ \ref{figure:Nsig cap}. We also find that $68.3\%$ of the values fall within $-1.55\leq N_{\sigma}\leq1.05$ and $95.4\%$ fall within $-3.66\leq N_{\sigma}\leq2.06$. For the absolute case, $|N_{\sigma}|\leq1.23$ and $|N_{\sigma}|\leq3.03$ for $68.3\%$ and $95.4\%$ of the values respectively. In terms of percentages, $61.0\%$ and $85.2\%$ of the measurements fall within $|N_{\sigma}|\leq1$ and $|N_{\sigma}|\leq2$ respectively. We note that when truncated, the normal standard deviation becomes $\sigma=0.125$ while the symmetrized error for the median case is $\sigma=0.126$. It would appear that after eliminating $|N_{\sigma}|>4$, the median and weighted mean cases converge. However, we do utilize a weighted mean rather than the standard mean, as the errors for the measurements are not the same, and the weighted mean and median statistics error still do not converge even in the truncated case. 
 
\begin{deluxetable}{lclccc} 
\tablecaption{Expected Gaussian and Observed Numbers of $|N_{\sigma}|$}
\tablewidth{0pt}
\tablehead{ 
\colhead{Tracer}& 
\colhead{Values\tablenotemark{a}}&
\colhead{$|N_{\sigma}|$}& 
\colhead{Expected\tablenotemark{b}}&
\colhead{Observed (WM)\tablenotemark{c}}&
\colhead{Observed (Med)\tablenotemark{c}}\\
}
\startdata
All Types  &   232   & $\geq0.5$&   143&	 151& 96\\
		&	      &	 $\geq1$&	     74&	 101&  45\\
		&	      &	 $\geq1.5$&   31&	 65&	   15\\
		&	      &	 $\geq2$&	     11&	 42&	  7\\
		&	      &	 $\geq2.5$&   3&	 31&	  1\\
		&	      &	 $\geq3$&	     1&	 23&	  0\\
		&	      &	 $\geq4$&	     0&	 9  &	  0\\
Cepheids  &    81   & $\geq0.5$&   50&	 48&   34\\
		&	      &	 $\geq1$&	     26&	 35&   17\\
		&	      &	 $\geq1.5$&  11&	 20&    4\\
		&	      &	 $\geq2$&	     4&	 10&   2\\
		&	      &	 $\geq2.5$&   1&	 7&     1\\
		&	     &	 $\geq3$&	     0&	 6  &	  0\\ 
RR Lyrae        &    63    & $\geq0.5$&   39&	 31&   20\\
		&	      &	 $\geq1$&	     20&	 20&   11\\
		&	      &	 $\geq1.5$&   8&	 12&   4\\
		&	     &	 $\geq2$&	     3&	 7  &	  1\\
		&	      &	 $\geq2.5$&  0&	 5&     0\\
		&	     &	 $\geq3$&	     0&	 3  &	  0\\						   
\noalign{\vskip 1mm}
\enddata
\tablenotetext{a}{The number of distance moduli measurements used in our analysis.}
\tablenotetext{b}{The number of values expected to fall outside of the corresponding $|N_{\sigma}|$ for a Gaussian distribution of total number listed in Col (2).}
\tablenotetext{c}{The observed number of values outside of the corresponding $|N_{\sigma}|$.}
\label{table:expected events Gaussian}
\end{deluxetable}
 
It is also of interest to determine the probabilities of the four well-known distributions for the new truncated weighted mean case. The distributions have a probability of $<0.1\%$ for the Gaussian, Cauchy, and double exponential distributions. The truncation to $|N_{\sigma}|<4$ improves the probability for the Student's $t$ case, slightly increasing it to $28\%$ compared to the $21\%$ for the non-truncated case. Since the probability does not improve for the Gaussian fit, this still indicates non-Gaussianity in the measurement distribution, because of the larger than expected $|N_{\sigma}|>$ 2 and 3 tail.

\section{Error Distributions for 81 Cepheid Values}
\label{Cepheids}

It is of interest to also investigate the spread of individual tracer measurements. We first consider the 81 Cepheid distance moduli values tabulated by \cite{Grijs2014}. For the weighted mean case we find a central estimate of  $(m-M)_{0}=18.52 \pm 6.52\times 10^{-3}$ mag.\footnote{We also find a $\chi^{2}=2.66$ and $N=7.96$ which is the number of standard deviations that $\chi$ deviates from unity.} For signed $N_{\sigma}$, $68.3\%$ of the values fall within $-1.04\leq N_{\sigma}\leq1.73$ and $95.4\%$ fall within $-1.78\leq N_{\sigma}\leq5.68$. For absolute $N_{\sigma}$, $68.3\%$ and $95.4\%$ of the values fall within $|N_{\sigma}|\leq1.31$ and $|N_{\sigma}|\leq4.13$ respectively, while $56.8\%$ of the values fall within $|N_{\sigma}|\leq1$ and $87.7\%$ fall within $|N_{\sigma}|\leq2$. 

For the median case we find a central estimate of $(m-M)_{0}= 18.50$ mag with a $1\sigma$ range of $18.37$ mag $\leq (m-M)_{0}\leq18.60$ mag. For signed $N_{\sigma}$, $68.3\%$ of the values fall within $-0.67\leq N_{\sigma}\leq0.73$ and $95.4\%$ fall within $-1.28\leq N_{\sigma}\leq1.76$. For absolute $N_{\sigma}$, $68.3\%$ and $95.4\%$ of the values fall within $|N_{\sigma}|\leq0.71$ and $|N_{\sigma}|\leq1.63$ respectively, while $79.0\%$ of the values fall within $|N_{\sigma}|\leq1$ and $97.5\%$ fall within $|N_{\sigma}|\leq2$. We note that for the median case, the error distribution is tighter when we use only the 81 Cepheid values compared to the distribution from all 232  measurements. 

\begin{center}
\begin{figure}[H]
\advance\leftskip-1.25cm
\advance\rightskip-1.25cm
\includegraphics[height=68mm,width=95mm]{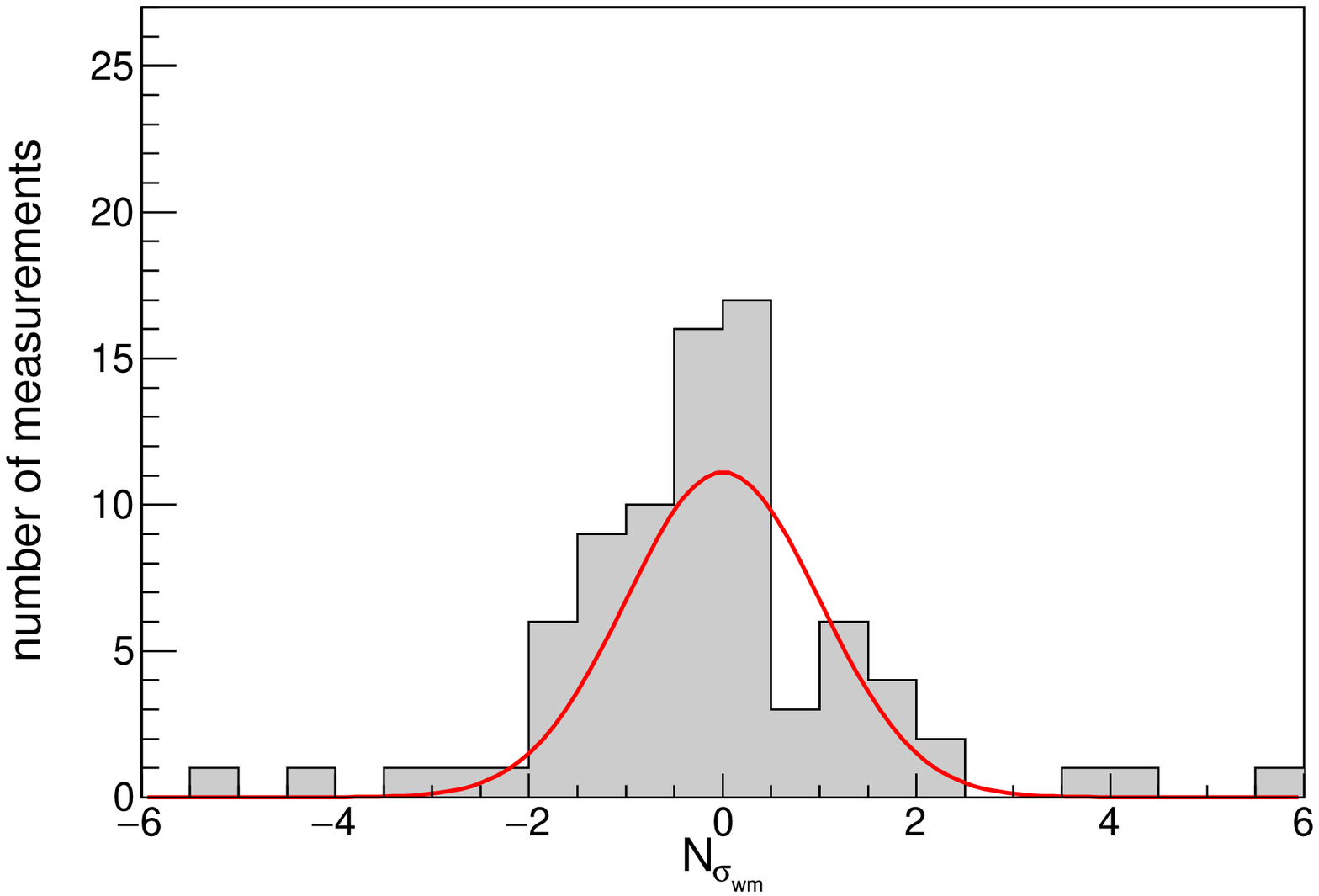}
\includegraphics[height=68mm,width=95mm]{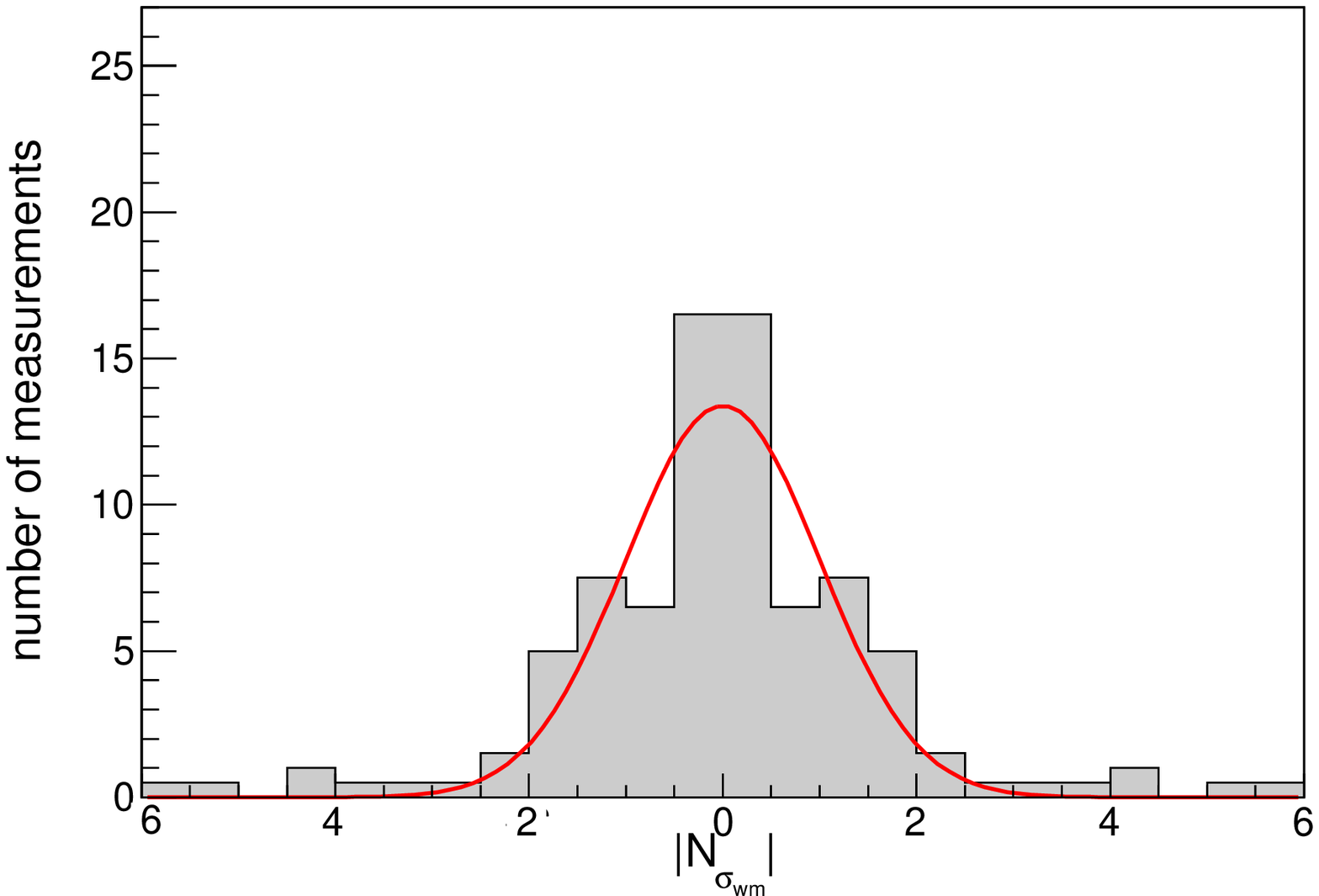}
\includegraphics[height=68mm,width=95mm]{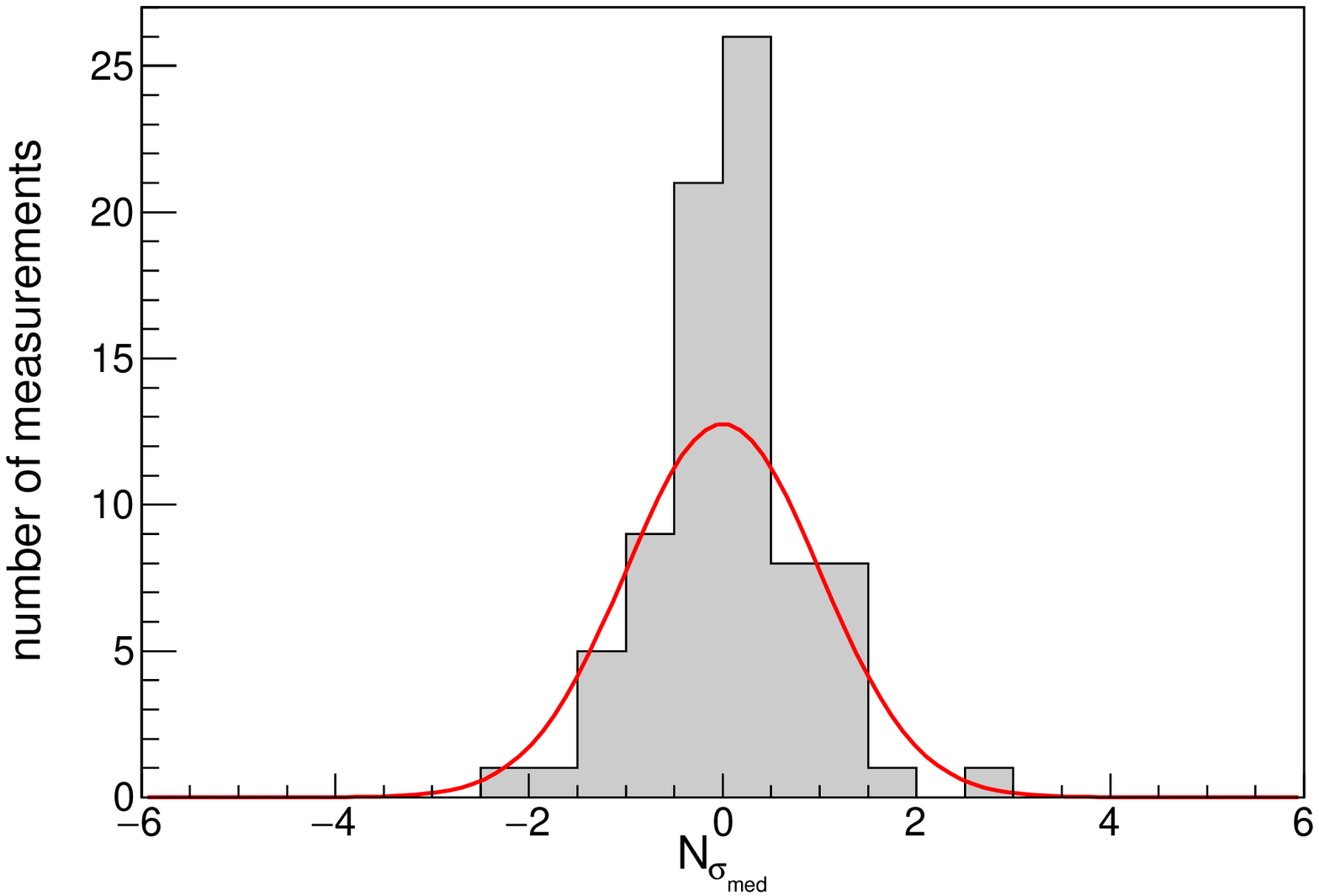}
\includegraphics[height=68mm,width=95mm]{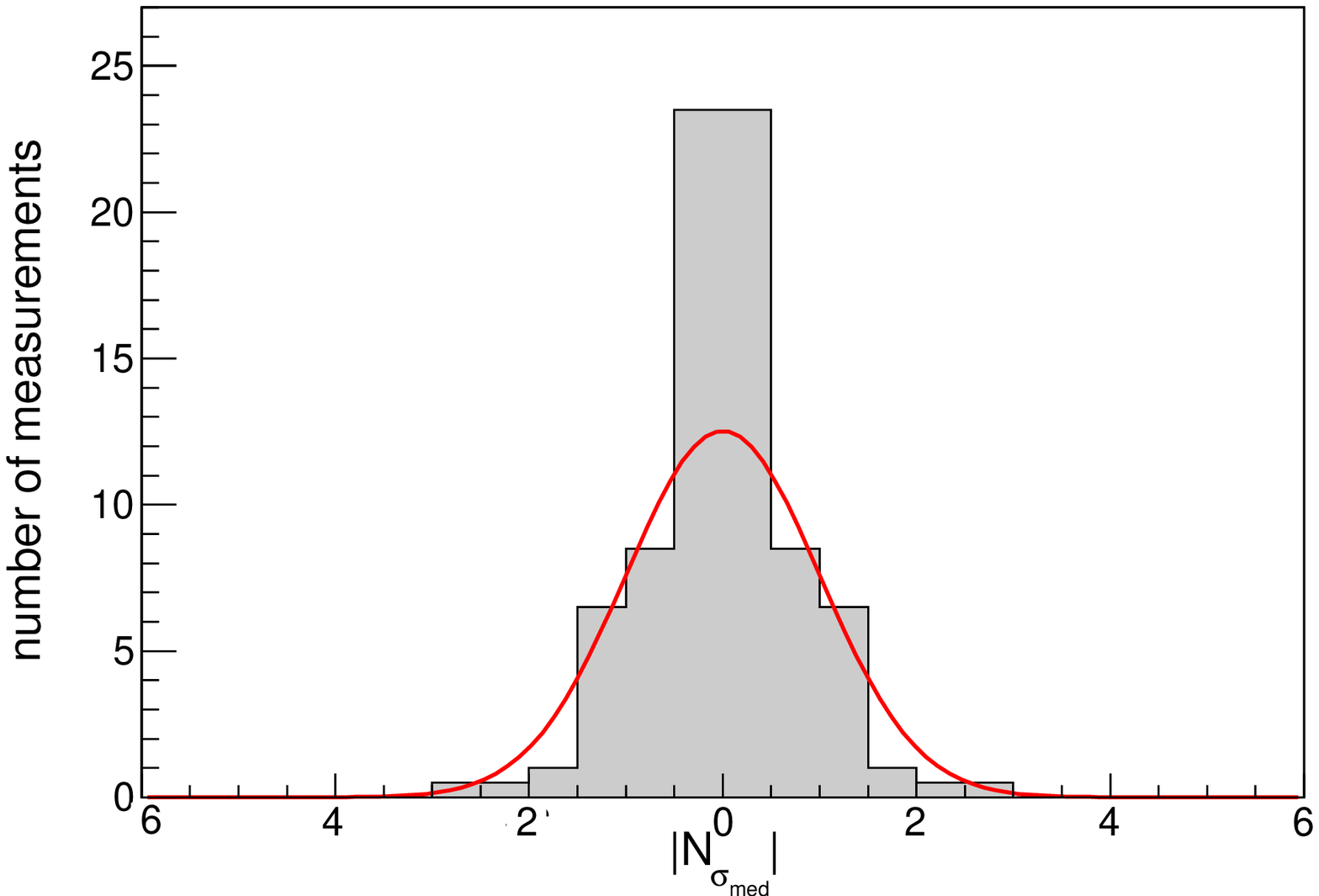}
\caption{Histograms of the error distribution in half standard deviation bins for Cepheids. The top (bottom) row uses the weighted mean (median) of the 81 measurements as the central estimate. The left column shows the signed deviation, where positive (negative) $N_{\sigma}$ represent a value that is greater (less) than the central estimate. The right column shows the absolute symmetrized distributions. The smooth curve in each panel is the best-fit Gaussian.} 
\label{figure:Nsig ceph}
\end{figure}
\end{center}

\begin{figure}[H]
\includegraphics[width=\linewidth]{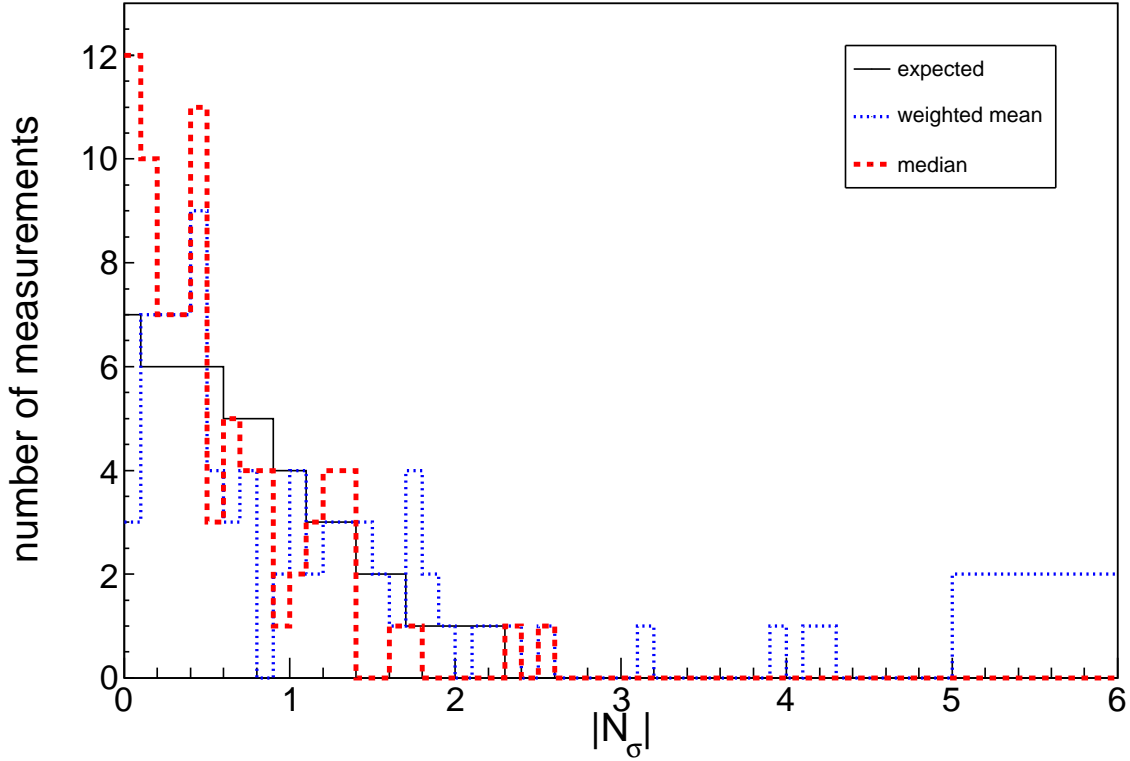}
\caption{Histogram of the error distributions using Cepheid tracers in $|N_{\sigma}|=0.1$ bins (with the exception of the last bin with $5\leq|N_{\sigma}|\leq6$). The solid black line represents the expected Gaussian probabilities for 81 measurements and the dotted blue (dashed red) line is the number of $|N_{\sigma}|$ values in each bin for the weighted mean (median) case.} 
\label{figure:prob ceph}
\end{figure}

The signed and absolute $N_{\sigma}$ distribution for the Cepheid tracers can be seen in Fig.\ \ref{figure:Nsig ceph}. One can see from the top two plots that there is an extended tail in the distribution for the weighted mean case, and from the lower two plots, a narrower distribution for the median case. We also plot $|N_{\sigma}|$ in bins of 0.1, see Fig.\ \ref{figure:prob ceph}. This figure again illustrates the higher than expected peak and rapid drop off of $|N_{\sigma}|$ for the median case, and the extended tails for the weighted mean case. To numerically describe these features, we can again use the four well-known distributions. The best probability comes from the Student's $t$ distribution for the median case with a probability of $26\%$ (see Table \ref{table:KS}). 

It is of interest to also truncate the Cepheid sub-sample, and we do so by truncating all $N_{\sigma}>3$ as there should be none for a normally distributed set of 81 measurements (see Table \ref{table:expected events Gaussian}). For the weighted mean case, with a new central estimate of $(m-M)_{0}=18.51 \pm 7.27\times 10^{-3}$ mag, the distribution slightly tightens. For signed $N_{\sigma}$, $68.3\%$ of the values fall within $-0.931\leq N_{\sigma}\leq1.48$ and $95.4\%$ fall within $-2.43\leq N_{\sigma}\leq2.38$. For absolute $N_{\sigma}$, $68.3\%$ and $95.4\%$ of the values fall within $|N_{\sigma}|\leq1.11$ and $|N_{\sigma}|\leq2.23$ respectively, while $65.3\%$ of the values fall within $|N_{\sigma}|\leq1$ and $94.7\%$ fall within $|N_{\sigma}|\leq2$. As for the median case, the distribution does not significantly change (as expected with median statistics). Table \ref{table:KS} shows the probabilities for the new truncated cepheid set. The probability for the un-binned KS test only slightly increases to $2.7\%$ while the binned probability does not significantly change.

\section{Error Distribution for 63 RR Lyrae Values}
\label{Lyrae}

\cite{Grijs2014} also tabulate 63 RR Lyrae distance moduli,\footnote{Three RR Lyrae values in \cite{Grijs2014} quote a zero error and were not used here.} whose error distribution we study here. We find a weighted mean central estimate of  $(m-M)_{0}=18.48 \pm 1.03\times 10^{-2}$ \textbf{mag}. For signed $N_{\sigma}$, $68.3\%$ of the values fall within $-0.83\leq N_{\sigma}\leq1.15$ and $95.4\%$ fall within $-1.75\leq N_{\sigma}\leq3.11$. For absolute $N_{\sigma}$, $68.3\%$ and $95.4\%$ of the values fall within $|N_{\sigma}|\leq1.00$ and $|N_{\sigma}|\leq3.11$ respectively, while $68.3\%$ of the values fall within $|N_{\sigma}|\leq1$ and $88.9\%$ fall within $|N_{\sigma}|\leq2$.  

For the median case we find a central estimate of $(m-M)_{0}= 18.47$ mag with a $1\sigma$ range of $18.29$ mag $\leq (m-M)_{0}\leq18.55$ mag. For signed $N_{\sigma}$, $68.3\%$ of the values fall within $-0.65\leq N_{\sigma}\leq0.48$ and $95.4\%$ fall within $-1.50\leq N_{\sigma}\leq1.03$. For absolute $N_{\sigma}$, $68.3\%$ and $95.4\%$ of the values fall within $|N_{\sigma}|\leq0.50$ and $|N_{\sigma}|\leq1.56$ respectively, while $82.5\%$ of the values fall within $|N_{\sigma}|\leq1$ and $98.4\%$ fall within $|N_{\sigma}|\leq2$.  

\begin{figure}[H]
\includegraphics[width=\linewidth]{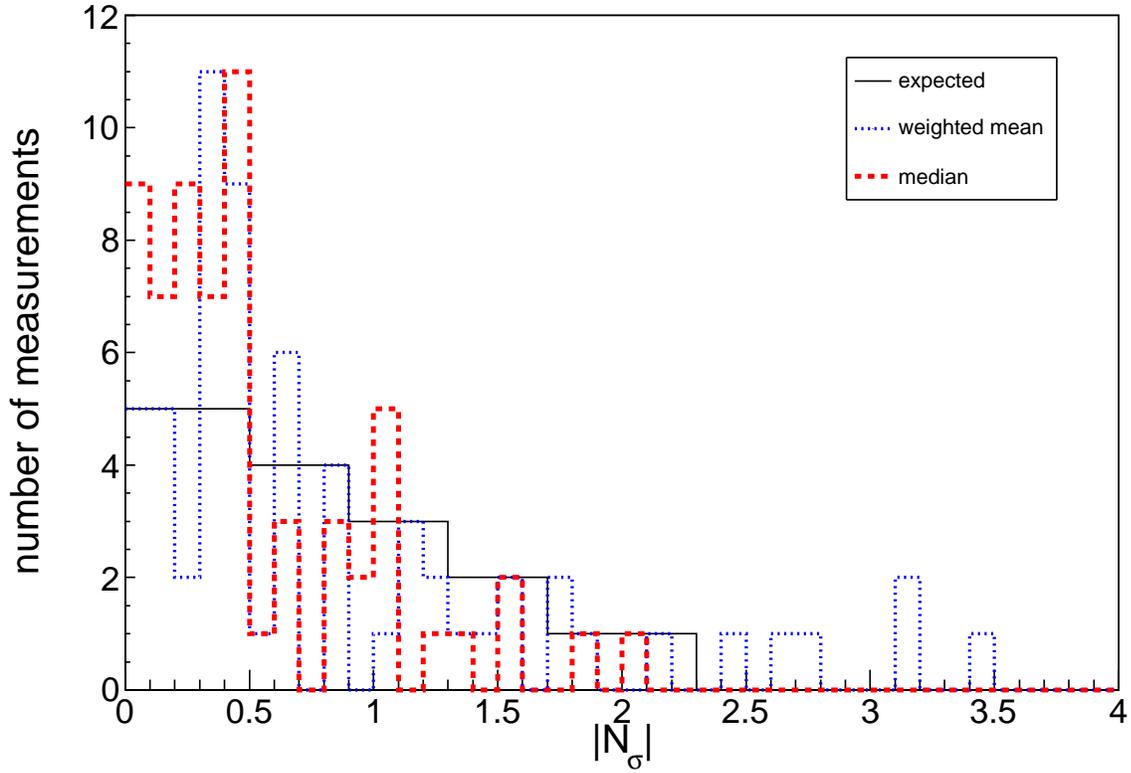}
\caption{Histogram of the error distributions using RR Lyrae tracers in $|N_{\sigma}|=0.1$ bins. The solid black line represents the expected Gaussian probabilities for 63 measurements and the dotted blue (dashed red) line is the number of $|N_{\sigma}|$ values in each bin for the weighted mean (median) case.}
\label{figure:prob lyrae}
\end{figure}

\begin{center}
\begin{figure}[H]
\advance\leftskip-1.25cm
\advance\rightskip-1.25cm
\includegraphics[height=68mm,width=95mm]{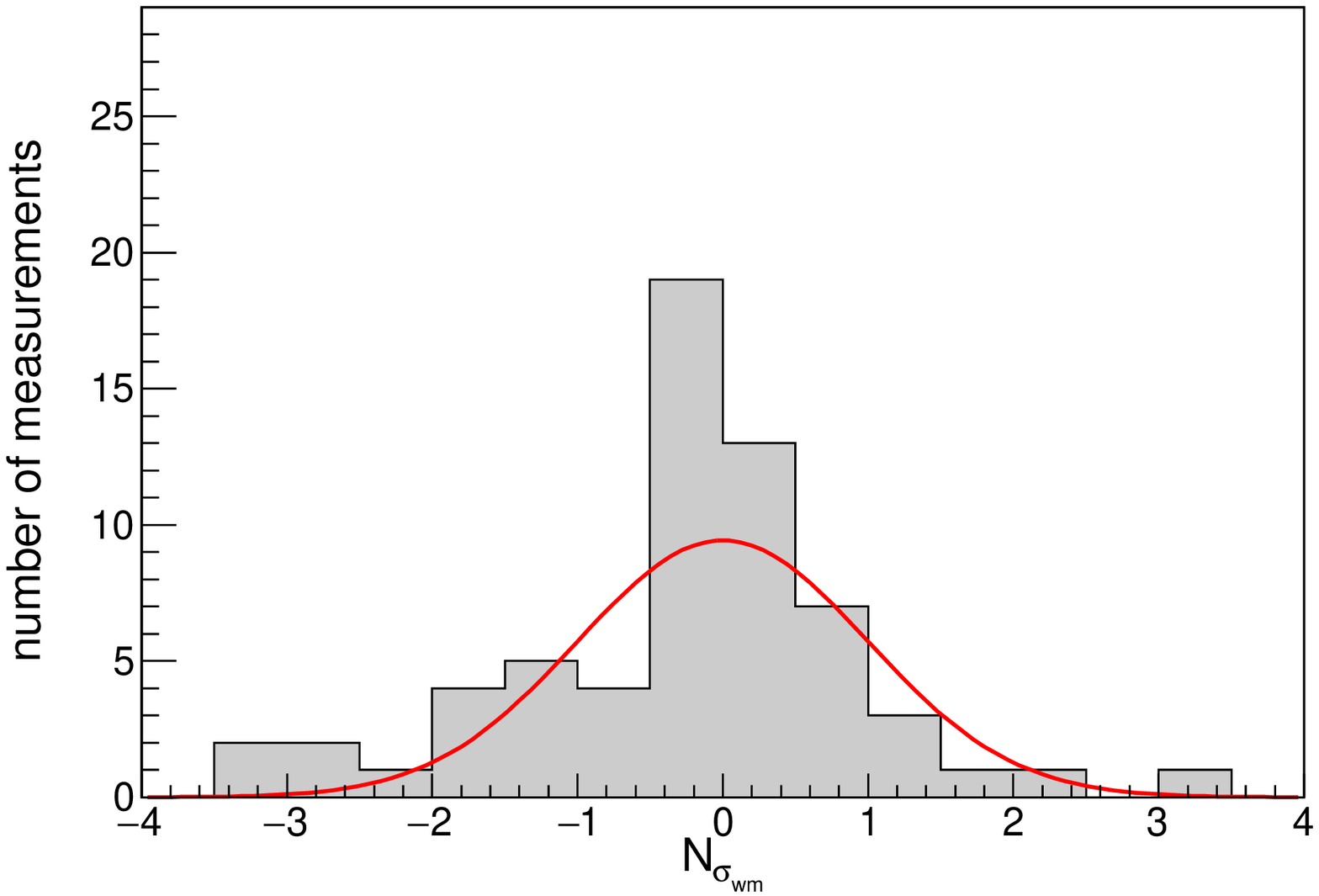}
\includegraphics[height=68mm,width=95mm]{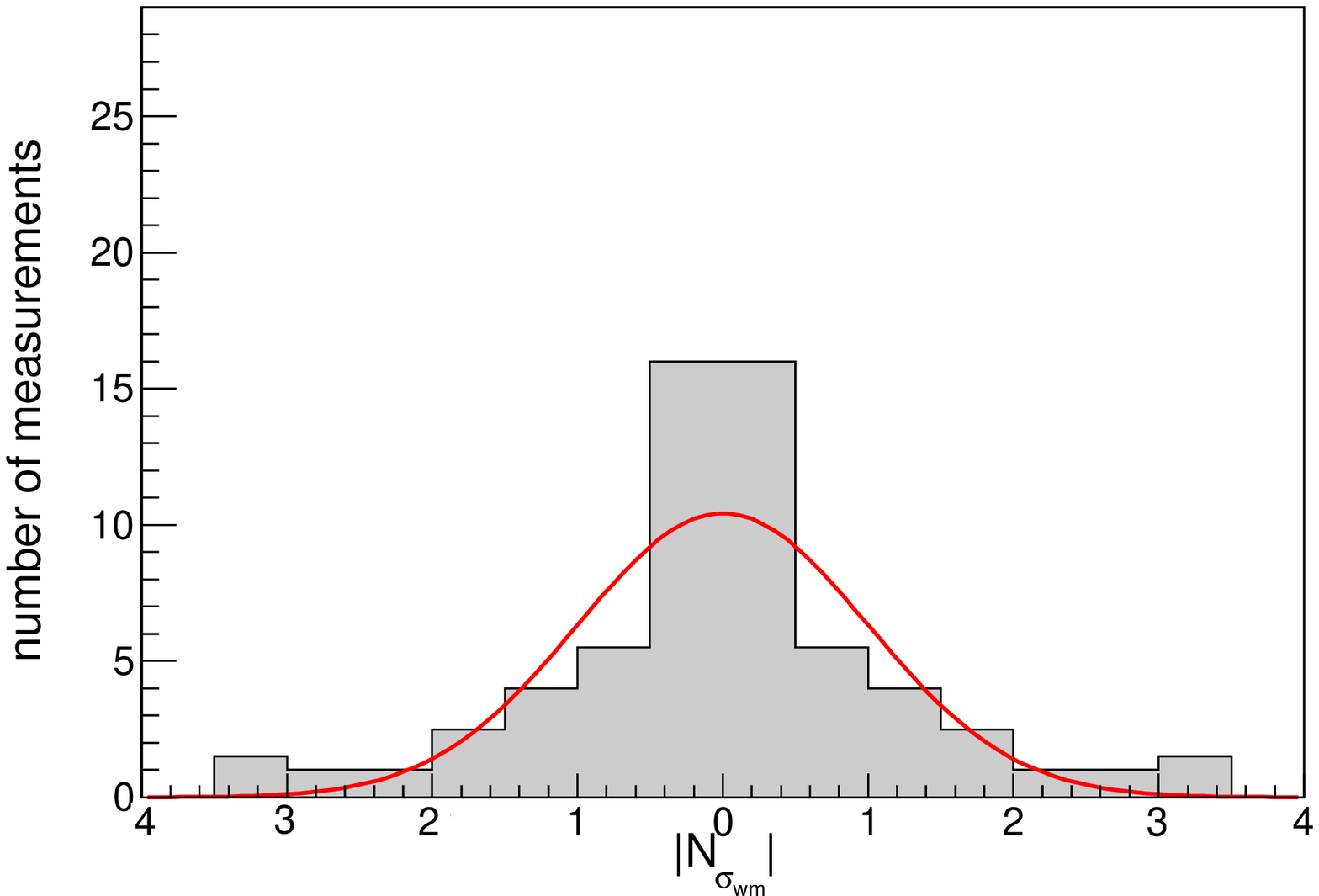}
\includegraphics[height=68mm,width=95mm]{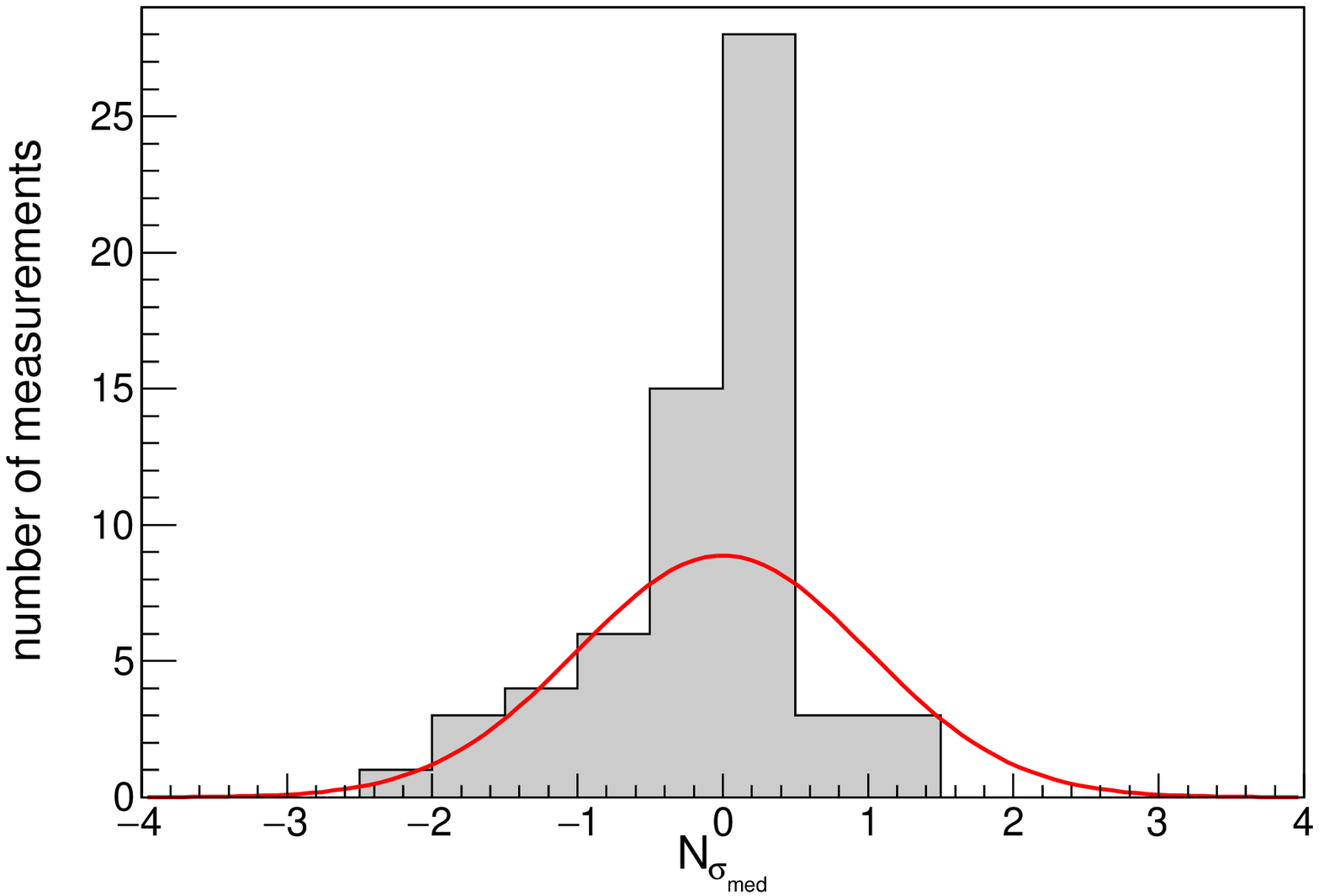}
\includegraphics[height=68mm,width=95mm]{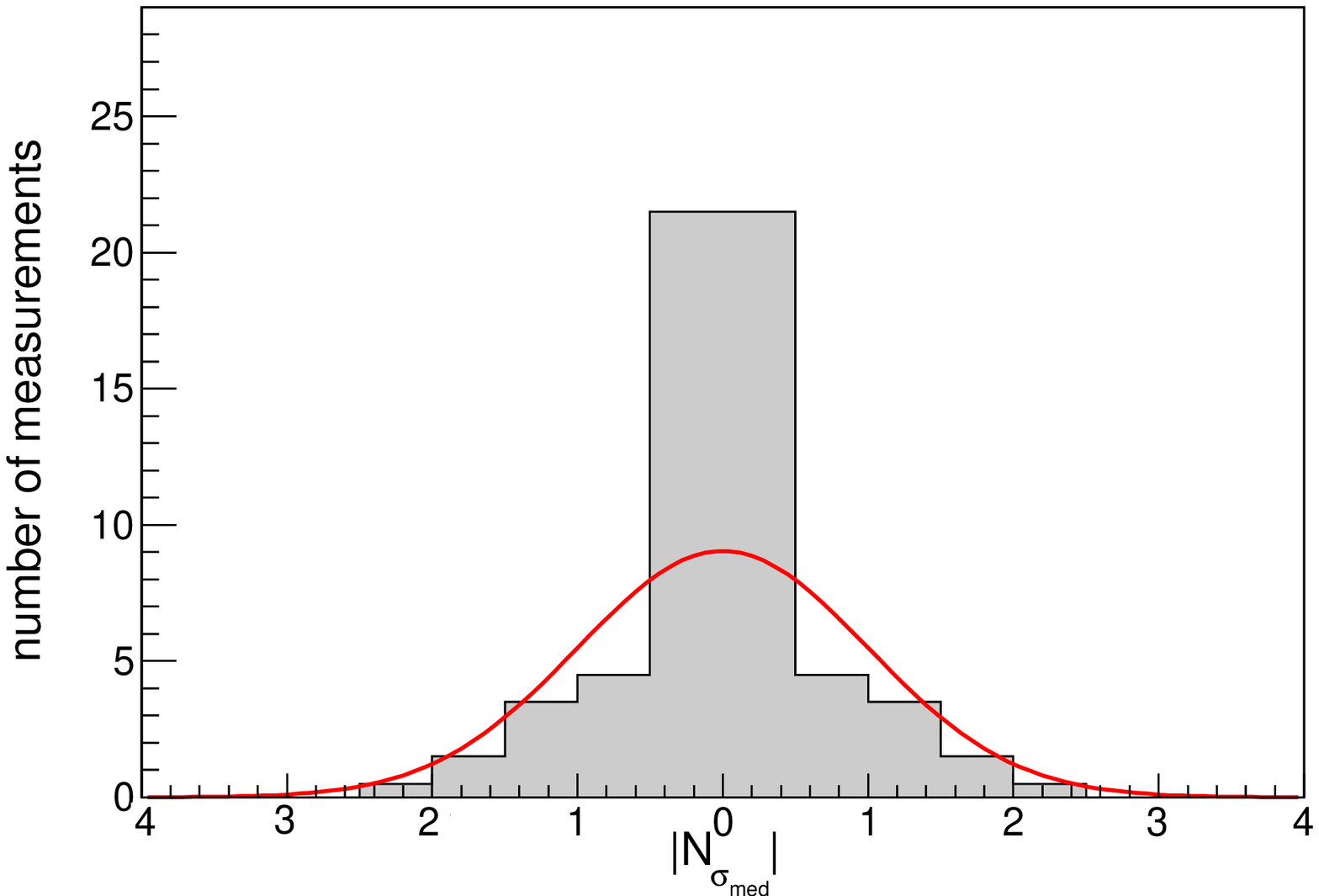}
\caption{Histograms of the error distribution in half standard deviation bins for RR Lyrae. The top (bottom) row uses the weighted mean (median) of the 63 measurements as the central estimate. The left column shows the signed deviation, where positive (negative) $N_{\sigma}$ represent a value that is greater (less) than the central estimate. The right column shows the absolute symmetrized distributions. The smooth curve in each panel is the best-fit Gaussian.} 
\label{figure:Nsig lyrae}
\end{figure}
\end{center}

We plot $|N_{\sigma}|$ in bins of 0.1, see Fig.\ \ref{figure:prob lyrae}, and the spread of values can be seen in Fig.\ \ref{figure:Nsig lyrae}. In this case, the non-Gaussianity is not as visually striking. We also fit the RR Lyrae measurements to the four distributions. The Student's $t$ distribution gives the largest probability of $34\%$(see Table \ref{table:KS}). 

We also truncated the RR Lyrae sub-sumple by only including values with $N_{\sigma}<2.5$, as there should be none greater than this for a set of 63 normally distributed measurements (See Table \ref{table:expected events Gaussian}.). In doing so the weighted mean error distribution was slightly tightened. \footnote{The median case did not significantly change.} We find a slightly changed central estimate of $(m-M)_{0}=18.49 \pm 1.10\times 10^{-2}$ mag. For signed $N_{\sigma}$, $68.3\%$ of the values fall within $-0.68\leq N_{\sigma}\leq0.79$ and $95.4\%$ fall within $-2.49\leq N_{\sigma}\leq1.81$. For absolute $N_{\sigma}$, $68.3\%$ and $95.4\%$ of the values fall within $|N_{\sigma}|\leq0.75$ and $|N_{\sigma}|\leq1.87$ respectively, while $75.9\%$ of the values fall within $|N_{\sigma}|\leq1$ and $98.3\%$ fall within $|N_{\sigma}|\leq2$. When fitting the sub-sample error distribution to well-known distributions, we find a slight increase in probabilities given by the KS test (See Table \ref{table:KS}.). We find that the highest probability of $37\%$ is given by an $n=94$ Student's $t$ distribution, which slightly increased from $34\%$.

\section{SMC Distance Moduli}
\label{SMC}
We have also analyzed 247 SMC distance moduli measurements\footnote{\cite{Grijs2015} collected 304 estimates, but we have only included measurements with non-zero error.} compiled by \cite{Grijs2015} and find similar results to those given by LMC distance Moduli measurements.\footnote{We thank Jacob Peyton for helping with this analysis.}  For the weighted mean case, which gives a central estimate of $(m-M)_{0}=18.93 \pm 2.38\times 10^{-2}$ mag, we find extended tails in the error distribution. For signed $N_{\sigma}$, we find that $68.3\%$ and $95.4\%$ of the measurements fall within $-2.01\leq N_{\sigma}\leq1.91$ and $-6.59\leq N_{\sigma}\leq4.76$ respectively. For the unsigned $N_{\sigma}$ we find that $68.3\%$ and $95.4\%$ of the measurements are within $|N_{\sigma}|\leq1.91$ and $|N_{\sigma}|\leq5.26$ respectively. Conversely, $45.8\%$ of the measurements fall within $|N_{\sigma}|\leq1$ and $70.5\%$ fall within $|N_{\sigma}|\leq2$. These wider tails suggest unaccounted-for systematic errors.

As for the median case, which gives a central estimate of $(m-M)_{0}=18.94$ mag with a $1\sigma$ range of $18.81$ mag $\leq (m-M)_{0}\leq19.08$ mag, the distribution is narrower than expected for a Gaussian. For signed $N_{\sigma}$, we find that $68.3\%$ and $95.4\%$ of the measurements fall within $-0.80\leq N_{\sigma}\leq0.78$ and $-1.60\leq N_{\sigma}\leq2.91$ respectively. For the unsigned $N_{\sigma}$ we find that $68.3\%$ and $95.4\%$ of the measurements are within $|N_{\sigma}|\leq0.79$ and $|N_{\sigma}|\leq1.68$ respectively. Conversely, $78.5\%$ of the measurements fall within $|N_{\sigma}|\leq1$ and $96.8\%$ fall within $|N_{\sigma}|\leq2$. This narrow distribution indicates the presence of correlations between measurements (especially within similar tracer types), as suggested by \cite{Grijs2015}.

We also examine the distributions given by the two tracer types with a greater number of measurements: Cepheids (101 measurements) and RR Lyrae (30). For the Cepheid weighted mean case, we find a central estimate of $(m-M)_{0}=18.98 \pm 4.17\times 10^{-3}$ mag. $68.3\%$ and $95.4\%$ of the measurements are within $-1.55\leq N_{\sigma} \leq 0.98$ and $-6.36\leq N_{\sigma} \leq 1.76$ for signed $N_{\sigma}$. For the absolute case $|N_{\sigma}|\leq1.26$ and $|N_{\sigma}|\leq4.02$ for $68.3\%$ and $95.4\%$ of the measurements respectively. Alternatively, $56.5\%$ and $86.1\%$ of the measurements fall within $N_{\sigma}\leq1$ and $N_{\sigma}\leq2$ respectively. For the median case, we find a central estimate of $(m-M)_{0}=18.98$ mag with a $1\sigma$ range of $18.83$ mag $\leq (m-M)_{0}\leq19.13$ mag. The distribution shows that $68.3\%$ and $95.4\%$ of the measurements are within $-0.83\leq N_{\sigma} \leq 0.68$ and $-1.91\leq N_{\sigma} \leq 1.46$ for signed $N_{\sigma}$. For the absolute case $|N_{\sigma}|\leq 0.81$ and $|N_{\sigma}|\leq 1.54$ for $68.3\%$ and $95.4\%$ of the measurements respectively. Alternatively, $82.2\%$ and $98.0\%$ of the measurements fall within $N_{\sigma}\leq1$ and $N_{\sigma}\leq2$ respectively. Again, we see a wider (narrower) than Gaussian distribution for the weighted mean (median) case. 

For the sub-sample of RR Lyrae tracer types, we notice similar distributions. For the weighted mean case, we find a central estimate of $(m-M)_{0}=18.86 \pm 5.20\times 10^{-3}$ mag. $68.3\%$ and $95.4\%$ of the measurements are within $-2.40\leq N_{\sigma} \leq 0.88$ and $-2.40\leq N_{\sigma} \leq 1.47$ for signed $N_{\sigma}$.\footnote{The two lower bounds are the same due to the distribution being weighted towards the positive $N_{\sigma}$ side (there are more values with $N_{\sigma}>0$). Symmetrizing this distribution gives a clearer understanding of the error.} For the absolute case $|N_{\sigma}|\leq1.49$ and $|N_{\sigma}|\leq3.26$ for $68.3\%$ and $95.4\%$ of the measurements respectively. Alternatively, $50.0\%$ and $80.0\%$ of the measurements fall within $N_{\sigma}\leq1$ and $N_{\sigma}\leq2$ respectively. For the median case, we find a central estimate of $(m-M)_{0}=18.90$ mag with a $1\sigma$ range of $18.74$ mag $\leq (m-M)_{0}\leq19.06$ mag. The distribution shows that $68.3\%$ and $95.4\%$ of the measurements are within $-0.51\leq N_{\sigma} \leq 0.86$ and $-1.13\leq N_{\sigma} \leq 1.24$ for signed $N_{\sigma}$. For the absolute case $|N_{\sigma}|\leq 0.65$ and $|N_{\sigma}|\leq 1.28$ for $68.3\%$ and $95.4\%$ of the measurements respectively. Alternatively, $83.3\%$ and $100\%$ of the measurements fall within $N_{\sigma}\leq1$ and $N_{\sigma}\leq2$ respectively.

We also attempt to fit the error distributions to four well-known distributions. The probabilities, found by using the KS test, are given in Table \ref{table:KSsmc}. We find that all distributions are fit best by a Student's $t$ distribution. The whole (247) distribution is best fit by an $n=1$ Student's $t$ with a probability of $74\%$.


\begin{deluxetable}{llcc} 
\tablecaption{K-S Test Probabilities}
\tablewidth{0pt}
\tabletypesize{\small}

\tablehead{
&&Un-binned&Binned\\
\colhead{Function\tablenotemark{a}}&
\colhead{Data Set}&
\colhead{Probability(\%)\tablenotemark{b}}&
\colhead{Probability(\%)\tablenotemark{b}}\\
\noalign{\vskip -6mm}
}
\startdata
\noalign{\vskip -0mm}
\noalign{\vskip 0mm}
Gaussian			&	Whole (247)&				$<0.1$&		$<0.1$\\
				&	Cepheids (101)&			$<0.1$&		$<0.1$\\
				&	RR Lyrae (30)&				8.4&			20\\
Cauchy			&	Whole (247)&				$<0.1$&		$<0.1$\\
				&	Cepheids (101)&			$<0.1$&		$<0.1$\\
				&	RR Lyrae (30)&				33&			15\\
Double Exponential 	&	Whole(247)&				$<0.1$&		$<0.1$\\
				&	Cepheids (101)&			$<0.1$&		$<0.1$\\
				&	RR Lyrae (30)&				36&			20\\
$n=1$ Student's $t$&	Whole (247)&				59&			$<0.1$\\
$n=1$ Student's $t$	&	Cepheids (101)&			74&			$<0.1$\\
$n=11$ Student's $t$&	RR Lyrae (30)&				22&			31\\	
\enddata
\tablenotetext{a}{For the Student's $t$ case, the $n$ corresponding with the best probability is displayed.}
\tablenotetext{b}{The probability that the data set is compatible with the assumed distribution.}
\label{table:KSsmc}
\end{deluxetable}

\section{Conclusion}
We have studied the error distributions of LMC distance moduli compiled by \cite{Grijs2014}. We find that the error distributions are non-Gaussian with extended tails when using a weighted mean central estimate, probably as a consequence of unaccounted-for systematic errors. In fact, only 53 of the 237 values tabulated by \cite{Grijs2014} have a non-zero systematic error. Because the weighted mean error distributions are non-Gaussian, it is more appropriate to use the median statistics error distribution. 

The median statistics error distributions are narrower than Gaussian, supporting the conclusion of \cite{Grijs2014}, who argue that this is a consequence of correlations between some of the measurements, with publication bias possibly also contributing mildly.

\label{Conclusion}

\acknowledgements
We thank R. de Grijs, J. Wicker, and G. Bono for providing us with the data. In addition, we thank R. de Grijs, J. Wicker, G. Horton-Smith, and A. Ivanov for valuable comments and advice. Finally, we thank Jacob Peyton for the SMC distance moduli analysis. This work was supported in part by DOE grant DEFG 03-99EP41093 and NSF grant AST-1109275.

\end{document}